# The joint distribution of pin-point plant cover data: a reparametrized Dirichlet - multinomial distribution

*Christian Damgaard, Department of Bioscience, Aarhus University, Vejlsøvej 25, 8600 Silkeborg, Denmark*

*cfd@bios.au.dk, +45-30183153*

**Abstract**

A reparametrized Dirichlet-multinomial distribution is introduced, and the covariance matrix, as well as, the algorithm for calculating the PDF for *n* species are provided. The distribution is suited for modelling the joint distribution of pin-point cover data of spatially aggregated plant species, and the parametrization ensures that the degree of spatially aggregation is modelled by a parameter and the mean cover of the species is modelled by the other parameters. This last property is convenient for using the distribution in Bayesian hierarchical models where the mean cover of the different species typically is modelled as latent variables. Computer code for Wolfram Mathematica and STAN is provided.

# Introduction

The statistical treatment of whole plant communities with multiple species has traditionally relied on ordination methods that use *ad hoc* distance measures, e.g. Euclidian or Bray-Curtis distances, to reduce the species by sample abundance matrix into a vector of distances between samples. However, there is an increasing interest in the statistically modelling of species abundance matrices (e.g. Clark *et al.*, 2014, Warton *et al.*, 2015, Ovaskainen *et al.*, 2017). Additionally, it has been suggested that in order to infer ecological processes from observed patterns of species abundance we need to investigate the covariance in species abundance (e.g. Brown *et al.*, 2011) and, if possible, to compare the observed covariance to the expected covariance when spatially aggregated plants associate randomly (Damgaard *et al.*, 2018).

When modelling multi-species pin-point plant cover data, it is important to account for the spatial aggregation that is typically found among plant species, and a reparametrized Dirichlet-multinomial distribution has been suggested for modelling such cover data (Damgaard, 2015, Damgaard & Irvine, 2019).

The Dirichlet distribution is the conjugate prior of the multinomial distribution, and the Dirichlet - multinomial distribution is a multivariate generalization of the beta -binomial distribution, which has been used with success previously to model pin-point cover data of spatially aggregated plant species (Damgaard, 2009, Damgaard, 2013).

Previously the reparametrized Dirichlet-multinomial distribution has only been used when the number of species groups were fixed to either four or five aggregated species groups ecosystems (Damgaard, 2015, Damgaard *et al.*, 2017, Damgaard, 2019), and then the general distribution properties and likelihood functions were determined for this *a priori* fixed number of species groups. However, it is interesting to be able to let e.g. phylogeny, traits, and the response to environmental factors determine the number and composition of species groups (Ovaskainen *et al.*, 2017), and in this case it is necessary to know the properties of the proposed reparametrized Dirichlet-multinomial distribution and be able to calculate the likelihood function distribution function in the general *n* species case.

The aim of this study is to investigate the properties of the proposed reparametrized Dirichlet-multinomial distribution in the general *n* species case. More specifically, a graphical interpretations of the parameter that account for the spatial aggregation will be provided, as well as an expressions of the covariance matrix and an algorithm for calculating the probability distribution function in the general *n* species case.

## Statistical model

The measurement of pin-point cover data is a binary event, where a pin in the pin-point frame either touches or does not touch a plant species. A discrete stochastic vector, $Y = (y_1, \ldots, y_n)$, may be defined as the number of pins in a pin-point frame that hit one of *n* plant species. Typically, a pin will hit more than one species, and the sum of the stochastic vector Y may be larger than the number of pins in a frame. Since the individual plant species are spatially aggregated, the stochastic vector *Y* is assumed to be Dirichlet - multinomial mixture distributed as

$$Y \sim Mn\left(\sum_n y_i, (p_1, \ldots, p_{n-1}, 1 - p_1 - \cdots - p_{n-1})\right) \tag{1},$$

$$\Lambda\ (p_1, \ldots, p_{n-1}) \sim Dir\left(\frac{q_1 - q_1\delta}{\delta}, \ldots, \frac{q_{n-1} - q_{n-1}\delta}{\delta}, \frac{1-\delta}{\delta} - \frac{q_1 - q_1\delta}{\delta} - \cdots - \frac{q_{n-1} - q_{n-1}\delta}{\delta}\right)$$

where $p_i$ model the cover of species *i*. The parameters $p_i$ are probabilities, and $q_i$ and $\delta$ is defined in the interval $]0, 1[$.

## Properties of the reparametrized Dirichlet-multinomial distribution

The rationale for using the above reparameterization of the Dirichlet distribution (eq. 1) is that it has the useful property that the mean cover of the species is a parameter in the model,

$$E(p_1, \ldots, p_{n-1}) = (q_1, \ldots, q_{n-1}) \tag{2},$$

and $\delta$ is a measure of the intra-plot correlation due to spatial aggregation of the plant species (Damgaard, 2015). At the limit when $\delta \to 0$, the Dirichlet - multinomial distribution degenerate to the multinomial distribution.

The marginal distribution of $p_i$ for the reparametrized Dirichlet distribution is

$$Beta\left(\frac{q_i - q_i\delta}{\delta}, \frac{1-\delta}{\delta} - \frac{q_i - q_i\delta}{\delta}\right) \tag{3}.$$

The $n \times n$ covariance matrix for the reparametrized Dirichlet distribution for *n* species is,

$$\begin{bmatrix} q_1(1-q_1)\delta & \cdots & -q_1 q_n \delta \\ \vdots & \ddots & \vdots \\ -q_n q_1 \delta & \cdots & q_n(1-q_n)\delta \end{bmatrix} \tag{4}.$$

Thus the effect of increasing the parameter $\delta$ is to increase the variation in cover relative to the multinomial distribution (Fig. 1), and this flexibility provided by the parameter $\delta$ is needed when modelling the cover distribution of spatially aggregated plant species (Damgaard, 2015).

Fig. 1. Probability distribution function of the Dirichlet distribution in the three species case, $(p_1, p_2) \sim Dir\left(\frac{q_1 - q_1 \delta}{\delta}, \frac{q_2 - q_2 \delta}{\delta}\right)$, with $q_1 = 0.3, q_2 = 0.4$ and variable $\delta$.

A: $\delta = 0.01$                               B: $\delta = 0.2$

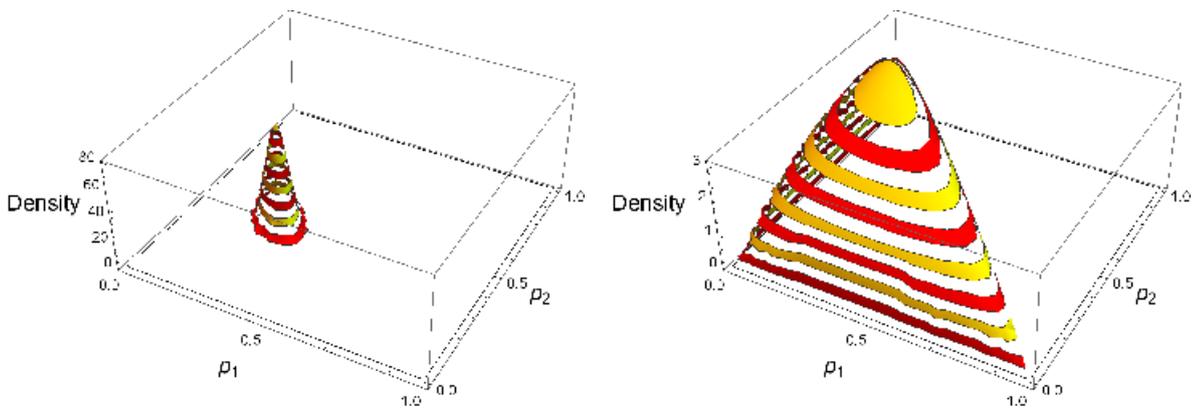

The Dirichlet - multinomial distributed model has been applied to model the vegetation dynamics in heathland ecosystems (Damgaard, 2015, Damgaard *et al.*, 2017, Damgaard, 2019).

## Algorithm for calculating the PDF

The PDF of the reparametrized Dirichlet - multinomial distribution for *n* species can be calculated using the following algorithm (written in the notation of a *Wolfram Mathematica* function),

pdfDM[n_, y_, q_, $\delta$_] := Module[{term1, term2, term3},

$$\text{term1} = \prod_{i=2}^{n} \text{Binomial}\left[y[[1]] + \sum_{j=2}^{i} y[[j]], y[[i]]\right];$$

$$\text{term2} = \left(\prod_{i=1}^{n-1} \text{Gamma}\left[y[[i]] + q[[i]]\left(-1 + \frac{1}{\delta}\right)\right]\right) * \text{Gamma}\left[-1 + \frac{1}{\delta}\right]$$

$$* \text{Gamma}\left[-1 + \left(\sum_{i=1}^{n-1} q[[i]]\right) + y[[n]] - \frac{-1 + \sum_{i=1}^{n-1} q[[i]]}{\delta}\right];$$

$$\text{term3} = \left(\prod_{i=1}^{n-1} \text{Gamma}\left[q[[i]]\left(-1 + \frac{1}{\delta}\right)\right]\right) * \text{Gamma}\left[-1 + \left(\sum_{i=1}^{n} y[[i]]\right) + \frac{1}{\delta}\right]$$

$$* \text{Gamma}\left[\frac{\left(-1 + \sum_{i=1}^{n-1} q[[i]]\right)(-1 + \delta)}{\delta}\right];$$

$$\text{Return}\left[\frac{\text{term1} * \text{term2}}{\text{term3}}\right]$$

]

The above algorithm for calculating the PDF of the Dirichlet - multinomial can easily be adopted into other programming languages, and the following example calculations is provided for check.

pdfDM[4, {4,4,1,6}, {0.2,0.3,0.1}, 0.1] = 0.00435623

pdfDM[5, {4,4,1,6,2}, {0.2,0.3,0.1,0.2}, 0.1] = 0.00045104230

See Appendix A for STAN code (Carpenter *et al.*, 2017).

Note that the above algorithm allow for inputs that is not a valid simplex, i.e., $\sum_{i=1}^{n-1} q[[i]] > 1$.

## When fitting a linear model to the mean cover of several species

When a linear model with several independent explaining variables is fitted to the mean cover of several species, then the linear model has to satisfy the constraints that $q_i > 0$ for all *i*, and $\sum q_i < 1$. In the single species case this is usually obtained by transforming the linear model with the inverse logit function. However, when the mean cover of several species are modelled it may be an advantage to modify the standard inverse logit transformation to $invlogit(x) = \frac{1/(n-1)}{1/(n-1)+\text{Exp}[-x]}$, so that $invlogit(0) = 1/n$ instead of 0.5.

For example, when $n = 4$, then $invlogit(x) = \frac{0.33}{0.33+\text{Exp}[-x]}$ may be used instead of the standard the inverse logit function. This will ensure that the above mentioned constraints on $q_i$ are met, when the parameters in the linear model is initialized to zero, e.g. in the beginning of an MCMC sampling procedure.

## Discussion

When modelling empirical plant abundance data it is critical to take the spatial aggregation, which is typically observed among plant species, into account (Damgaard & Irvine, 2019), and in the suggested reparametrized Dirichlet - multinomial distribution this is done by the parameter $\delta$, which increases with the within-site spatial variation in cover, i.e. the degree of spatial aggregation.

The parameters $q_i$ in the suggested reparametrized Dirichlet - multinomial distribution model the mean cover of species *i*. This is a convenient property when using hierarchical Bayesian models, where the mean cover of the different species typically is modelled as latent variables (Damgaard, 2015, Damgaard *et al.*, 2017, Damgaard, 2019). Furthermore, the Dirichlet-multinomial distribution is the marginal joint posterior distribution of multinomial distributed categorical observations, where the multinomial probability parameters are integrated out, which makes it possible to construct relatively fast MCMC sampler, e.g. by using Gibbs sampling.

The proposed parameterization of the Dirichlet distribution may be regarded as a null model for patchy and independently distributed plant species. However, it is possible to loosen the assumption of independence if plant abundance is modelled by common covariates, in which case within-plot positive covariation among species may be modelled (Hijazi & Jernigan, 2009).

Generally, it is expected that the Dirichlet-multinomial mixture distribution is a fairly good description of the joint distribution of the pin-point cover of a few or a moderate number of species. However, when the number of species is high it is expected that more than a single spatial aggregation parameter is needed to adequately model the joint cover distribution.

The presented results for the general *n* species case will enable the use of the reparametrized Dirichlet-multinomial distribution in the interesting case when the number of species groups are determined by the phylogeny, traits, and the response to environmental factors of the observed plant communities (Ovaskainen *et al.*, 2017). Furthermore, the statistical model may be used in model based ordination studies of plant cover pin-point data as outlined by Hui et al. (2015).

It is the hope that this short account will facilitate the application of the reparametrized Dirichlet-multinomial distribution.

## Acknowledgements

Thank you to Kathi Irvine and Hans van Calster for valuable discussions.

# Appendix A: algorithm for calculating the log PDF in STAN

```
functions{
 real dirichlet_multinomial_lpmf(int[] y, int S, real[] mu, real delta) {
   real term1;
   real term2;
   real term3;
   real sumvar;
   term1=1;
   for (i in 2:S){
     sumvar=0;
              for (j in 2:i)
                sumvar += y[j];
              term1 *= exp(lchoose(y[1] + sumvar, y[i]));
   }

   term2=1;
   for (i in 1:S-1)
      term2 *= tgamma(y[i]+mu[i]*(-1+1/delta));
   term2 *= tgamma(-1+1/delta);
   term2 *= tgamma(-1+sum(mu)+y[S]-(-1+sum(mu))/delta);

   term3=1;
   for (i in 1:S-1)
      term3 *= tgamma(mu[i]*(-1+1/delta));
   term3 *= tgamma(-1+sum(y)+1/delta);
   term3 *= tgamma((-1+sum(mu))*(-1+delta)/delta);

   return log(term1)+log(term2)-log(term3);
 }
}
```